\documentclass[fleqn,usenatbib]{mnras}

\usepackage{newtxtext,newtxmath}
\usepackage{tabularx}
\usepackage{tikz}
\usepackage{booktabs, caption, makecell}
\usepackage{subcaption}
\usepackage{multirow}
\usepackage[integrals]{wasysym} %
\usepackage{todonotes}

\usepackage[T1]{fontenc}
\usepackage{amsmath}	

\DeclareRobustCommand{\VAN}[3]{#2}
\let\VANthebibliography\thebibliography
\def\thebibliography{\DeclareRobustCommand{\VAN}[3]{##3}\VANthebibliography}

\usepackage{graphicx}	
\usepackage{amsmath}	
\usepackage{tikz}

\title[Radio Activity from 2MASS 2228-4310]{Radio Activity from the Rapidly Rotating T dwarf 2MASS 2228-4310}

\author[Kelvin Wandia et al.]{Kelvin Wandia,$^{1}$\thanks{E-mail: kelvin.wandia@manchester.ac.uk}
Michael A. Garrett,$^{1,2}$
Aaron Golden, $^{3}$
Gregg Hallinan, $^{4}$
David Williams-Baldwin,$^{1}$
\newauthor
Geferson Lucatelli, $^{1}$
Robert J. Beswick,$^{1}$
Jack F. Radcliffe,$^{1,5}$
Andrew Siemion,$^{1,6,7,8,9,10,11}$ 
\newauthor
Talon Myburgh $^{12}$
\\
$^{1}$Jodrell Bank Centre for Astrophysics (JBCA), Department of Physics \& Astronomy, Alan Turing Building, The University of Manchester, M13 9PL, UK\\
$^{2}$Leiden Observatory, Leiden University, PO Box 9513, 2300 RA Leiden, The Netherlands\\
$^{3}$Physics, School of Natural Sciences \& Center for Astronomy, College of Science and Engineering, University of Galway, University \\
Road, Galway, H91 TK33, Ireland \\
$^{4}$Division of Physics, Mathematics, and Astronomy, California Institute of Technology, Pasadena, CA 91125, USA \\
$^{5}$Department of Physics, University of Pretoria, Lynnwood Road, Hatfield, Pretoria, 0083, South Africa \\
$^{6}$SETI Institute, 339 Bernardo Ave, Suite 200, Mountain View, CA 94043, USA\\
$^{7}$Berkeley SETI Research Center, University of California, Berkeley, CA 94720, USA \\
$^{8}$Breakthrough Listen, Astrophysics, Department of Physics, The University of Oxford, Keble Road, Oxford, OX1 3RH, UK \\
$^{9}$Astrophysics, Department of Physics, University of Oxford, Keble Road, Oxford, OX1 3RH, UK \\
$^{10}$Berkeley SETI Research Center, University of California, Berkeley, CA 94720, USA \\
$^{11}$University of Malta, Institute of Space Sciences and Astronomy, Msida, MSD2080, Malta \\
$^{12}$ Centre for Radio Astronomy Techniques and Technologies, Department of Physics and Electronics, Rhodes University, Makhanda, 6140, 
South Africa
}

\date{Accepted XXX. Received YYY; in original form ZZZ}

\pubyear{\the\year{}}

\begin{document}
\label{firstpage}
\pagerange{\pageref{firstpage}--\pageref{lastpage}}
\maketitle

\begin{abstract}
We present the detection of 2MASS J22282889-4310262 (2M2228), a T6/T6.5 brown dwarf, using the Karl G. Jansky Very Large Array (VLA) archival data observed at C band (4\mbox{--}8 GHz) over two observing epochs ($2\times96$ minutes). 2M2228 is detected at time and frequency averaged Stokes I and V peak flux densities of $\textcolor{black}{67.3\pm4.9\ \mu\ \rm{Jy beam}^{-1}}$ and $\textcolor{black}{14.4\pm3.0\ \mu\text{Jy beam}^{-1}}$ in the first epoch and $\textcolor{black}{107.2\pm5.2\ \mu\rm{Jy\ beam}^{-1}}$ and $\textcolor{black}{-20.7\pm1.2\ \mu\text{Jy beam}^{-1}}$ in the second epoch.
This discovery constitutes the eighth and, notably, the most rapidly rotating T dwarf detected to date at radio wavelengths. Our observations reveal highly polarised bursts at fractional polarisation ratios $f_\text{c}>50$\%. Using Stokes I light curves, we measure occurrence intervals of $\sim47$ and $\sim58$ minutes in the two observing epochs respectively with the first burst aligning within a half period timescale of the the previously measured mid infrared photometric period of $85.8\pm0.32$ minutes.  We attribute the emission to the electron cyclotron maser emission (ECME) and constrain the magnetic field strength to $B\gtrsim1.4$ kG. We emphasise that the periods inferred are provisional considering the short observing durations. The combination of previously demonstrated atmospheric stability and newly detected radio emission in 2M2228 makes it a promising laboratory for testing magnetospheric currents-driven auroral models and for guiding future coordinated James Webb Space Telescope (JWST) and radio observations to probe the link between auroral activity and atmospheric dynamics in T-type brown dwarfs.



\end{abstract}




\section{Introduction}

After the first undisputed discovery of Gliese 229B, a brown dwarf of spectral type T \citep[][]{Nakajima1995,Oppenheimer1995}, the catalogue of brown dwarfs was substantially enhanced using the 2-Micron All-Sky Survey \citep[2MASS; ][]{Skrutskie2006,Kirkpatrick1999,Kirkpatrick2000} largely due to improvement in infrared telescopic equipment. This development prompted the definition of new classes of substellar objects denoted L and T dwarfs at effective temperatures ranging between  $T_\text{eff}\sim2500$--$1300$~K for L dwarfs and $T_\text{eff}\sim1500$--$750$~K  for T dwarfs \citep[][]{Kirkpatrick_2005_L_T_dwarfs}. Following the launch of the Wide-field Infrared Survey Explorer (WISE) telescope in 2009 and a subsequent full sky survey \citep[][]{Wright2010} a new class of substellar objects at $T_\text{eff}\lesssim500$ K \citep[][]{Cushing2011} were detected and classified as Y dwarfs. \textcolor{black}{These discoveries expanded our understanding of the low-temperature end of the substellar population, including} brown dwarfs which are intermediate objects between planets and stars. Although no sharp boundary exists between giant planets and brown dwarfs, a common criterion for separating the two is the minimum mass required for deuterium fusion, with objects roughly above 13 Jupiter masses classified as brown dwarfs \citep[][]{Spiegel2011}. The situation becomes more nuanced with the discovery of \textcolor{black}{planetary mass objects} with masses below \textcolor{black}{the deuterium burning limit} e.g. the free floating \textcolor{black}{object} in the star-forming cluster IC 348 at $3$--$4\ M_\text{J}$  \citep[][]{Luhman2024}. \textcolor{black}{Because they cannot fuse deuterium, these objects are not considered brown dwarfs, although they share other physical and spectral properties with low-mass brown dwarfs.} A new spectral class ``H'' has now been proposed to account for such \textcolor{black}{objects} \citep[][]{Luhman2025}. Conversely, brown dwarfs are separated from stars based on the mass limit of $\sim0.075~\mathrm{M}_{\odot}\ (78.5\ \mathrm{M}_\text{J})$  required to ignite and sustain hydrogen fusion \citep[][]{Chabrier2023}. \textcolor{black}{While all T and Y dwarfs are generally brown dwarfs, young gas giant exoplanets can temporarily reach similar temperatures. Conversely, some early type L dwarfs are very low-mass stars, and some late type M dwarfs can be brown dwarfs. Objects of spectral type >L2.5 are generally believed to be brown dwarfs \citep[][]{Dieterich2014}.} Brown dwarfs together with very low-mass stars of spectral type later that $\sim$\,M7V $(T_\text{eff}\lesssim2700\ \text{K})$ are collectively known as ultra cool dwarfs \citep[UCDs; ][]{Cushing2006} and are expected to be fully convective.

\textcolor{black}{Due to their convective interiors, UCDs are now known to be capable of supporting strong magnetic fields} which have been attributed to an $\alpha^2$ type dynamo at work in their interior \citep{Chabrier2006,Dobler2006,Browning2008}. \textcolor{black}{These magnetic fields drive activity that may be observed at the surfaces of earlier-type UCDs, but is largely absent in later types due to their increasingly neutral atmospheres. In addition to their magnetic properties, the cooler temperatures of later-type UCDs allow their atmospheres to form and sustain complex molecules}. Observations of their atmospheres in the mid infrared show photometric variabilities modulated by the rotation period caused by atmospheric heterogeneities and suggest the presence of complex weather patterns \cite[e.g.][]{Metchev_2015_II,Buenzli2012,Yang2016}.

One of the mechanisms through which UCDs generate radio emission is through the electron cyclotron maser emission \citep[ECME; ][]{Wu1979, Dulk1985, Melrose1982, Callingham2024}. ECME is a coherent emission mechanism arising from \textcolor{black}{non-thermal anisotropic electrons in a horseshoe or a loss cone distribution} \citep[][]{Melrose2017}. \textcolor{black}{As a coherent emission mechanims, ECME can generate extremely high brightness temperatures which can exceed the $10^{12}$ K inverse Compton limit that applies to incoherent synchrotron emission \citep[][]{Kellerman1969}. The resulting radio emission is often} highly circularly polarised emission, \textcolor{black}{frequently approaching 100\%, with detections of both helicities interpreted as arising from} magnetic distinct regions and or changing magnetic field orientations with respect to the viewing angle \citep[][]{Williams2018}. The radio emission is also often rotationally modulated, reflecting the periodic visibility of the emitting regions \citep[][]{Hallinan2006, Williams2018}. The mechanism driving coherent radio emission in UCDs is believed to be analogous to that of the Jovian system involving large scale magnetospheric currents \citep[][]{Hallinan2015} caused by co-rotation breakdown primarily due to mass loading either from orbiting planets/moons or the interstellar medium \citep[][]{Nichols2012}. \citet[][]{Hallinan2015} have also proposed that magnetospheric currents are likely to contribute to observed photometric variability that traces weather in brown dwarfs.

Although radio activity is now beginning to be characterised in UCDs, the first detection of a brown dwarf at radio wavelengths by \citet[][]{Berger2001} challenged standard theories of dynamo processes, activity, and temperature relationships. According to the temperature activity relationship, a hot chromosphere at a temperature of $\sim10^{4}$ K and the coronal at a temperature of $\sim10^{7}$ K emit H$\alpha$ and soft X-rays, respectively. The two act as tracers of magnetic activity. As such, the detection of strong magnetic fields in brown dwarfs, which do not appear to have a classic chromosphere and corona, came as a surprise. In addition, this detection violated the empirically derived Gudel-Benz relationship, which relates the X-ray fluxes from the corona to the 5 GHz radio emission \citep[][]{GudelBenz1993}. Subsequent detections of UCDs have been made at radio wavelengths \citep[e.g.][]{Berger2002,Berger2006,Berger2009,Burgasser2005,Burgasser2013,Hallinan2007,Hallinan2008,Hallinan2015,Mclean2011,McLean2012,Gizis2013,Gizis2016,Lynch2016, Vendatham2020b,Rose2023} and have expanded our understanding of magnetic/radio activity in low-mass stars and substellar objects. 

The coolest brown dwarfs detected at radio wavelengths are of spectral type T with seven confirmed detections.  They include the T6/T6.5  WISEP J112254.73+255021.5 \citep[][]{Route2016,Williams2017}, T6/T6.5.5 2MASS 10475385+2124234 \citep[][]{Route2012,Williams2013,Williams2015}, T2.5 SIMP J01365663+0933473 and T6/T6.5.5 2MASS J12373919+6526148 \citep[][]{Kao2016,Kao2018}, T6/T6.5 BDR J1750+3809 \citep[][]{Vendatham2020}, T7.0+T5.5 binary WISEP J101905.63+652954.2 \citep[][]{Vendatham2023} and the T8 WISE J062309.94$-$045624.6 \citep[][]{Rose2023}. \textcolor{black}{These detections reveal that ultra-cool, low mass objects can host strong magnetic fields and support auroral processes, highlighting a continuum between stellar and planetary magnetism. While radio
emission has not yet been detected from the coolest brown dwarfs, the Y dwarfs, observational constrains have been placed providing important limits on their magnetic activity \citep[][]{Kao2019}.} We note that the ultimate goal is the detection of magnetic fields from exoplanets, a prospect that may be realised with the upgraded Low Frequency Array (LOFAR 2.0) and the low frequency component of the Square Kilometre Array (SKA-low).

In response to the paucity of T dwarfs detected at radio wavelengths, we conducted a search of archival National Science Foundation Karl G. Jansky Very Large Array (VLA) data sets for observations targeting T-type brown dwarfs. We discovered unpublished (to the best of our knowledge) observations of the T6/T6.5 brown dwarf 2MASS J22282889-4310262. These observations aimed to extend the sample of the coldest brown dwarfs detected at radio wavelengths. In this manuscript, we describe the targets in Section 2, in Section 3 we outline the methodology used for the data analysis, in Section 4 we present our results and discussion, and finally, we summarise our findings in Section 5.


\section{2MASS J22282889-4310262}

The brown dwarf 2MASS J22282889-4310262 (hereinafter 2M2228) was first reported by \cite{Burgasser2003} and assigned the spectral class T6.5 but further re-classified as a T6 \citep[][]{Burgasser2006} and is at a distance of $\sim10.64\pm0.79$ pc \citep[][]{Faherty2012}. We note that both spectral types have been cited in the literature, depending on the reference used. 2M2228 is rapidly rotating at a photometric period of $1.43\pm0.16$ \textcolor{black}{hours} \citep[][]{Clarke2008}. The rotation period makes it \textcolor{black}{one of the most} rapidly rotating known UCD of spectral type T \citep[][]{Tannock2021,Vos2022}. 

2M2228 has a peak-to-peak photometric variability of $15.4\pm1.4$ milli-magnitude (mmag) in the J-band $(1.0$--$1.25\ \mu\text{m})$ \citep[][]{Clarke2008}. \textcolor{black}{ The atmosphere of the brown dwarf has been characterised using the Hubble Space Telescope (HST) Wide Field Camera 3 (WFC3) at $1.1$–$1.7\ \mu\mathrm{m}$ and the Spitzer Infrared Array Camera (IRAC) at $4.5\ \mu\mathrm{m}$ \citep{Buenzli2012}, which revealed vertical heterogeneities inferred from the phase shifts of light curves observed at the two wavelengths. Follow-up observations by \cite{Yang2016} found that the phase shifts between HST and Spitzer reported four years earlier by \cite{Buenzli2012} persisted, despite the measurements being separated by thousands of rotations, highlighting that the dwarf possesses long-lived vertical atmospheric structures. } .

\begin{table}
  \centering
    \renewcommand{\arraystretch}{1.2}
    \begin{tabular}{ccc} \\
		\hline
		 Property &  Value  & Reference   \\
         \hline
         Spectral type & T6/T6.5  & $1$ \\
         Distance (pc) & $ 10.64\pm0.79$ &  $2$ \\
         $T_\text{eff}$ (K) & 900 & $3$ \\
         J-band amplitude (mmag) & $15.4\pm1.4$ & $4$ \\
         Period (hrs) & $1.43\pm0.16$ & $4$ \\
         Radius $(R_\text{J})$ & $0.94\pm0.16$ & $5$  \\
         Age (Myr) & 1000 & $5$ \\ 
         \hline
	\end{tabular}
    \caption{Physical parameters References:$^1$\protect\cite{Burgasser2003,Burgasser2006}, $^2$\protect\cite{Faherty2012}, $^3$\protect\cite{Buenzli2012}, $^4$\protect\cite{Clarke2008}, $^5$\protect\cite{Vos2020},
    }
    \label{table:physical_params}
\end{table}

\section{Methods}

\subsection{Observations}

We used \textcolor{black}{archival} observations (project code 15A-045, PI Aaron Golden) of 2M2228 conducted between 29-05-2015 and 31-05-2015 at C band (4--8 GHz) over two separate epochs on the aforementioned dates. The low declination of the source ($\sim-43\degr$) required observations using the hybrid BnA configuration. The first epoch spanned 2 hours from 11:31 UTC to 13:31 UTC, and the second spanned a similar time frame from 11:16 UTC to 13:16 UTC, with the target observation beginning at 11:47 UTC for the first epoch and at 11:33 UTC for the second. Three sources were observed, 3C48 as the flux density calibrator, J2257-3657 as the phase calibrator, and 2M2228. The on-target time for each epoch was $\sim96$ minutes, yielding a total time of $\sim192$ minutes.
The data from the observations were processed using the \texttt{WIDAR} correlator and recorded at 2 seconds of integration and 64 channels over 32 spectral windows (spws) \textcolor{black}{each 128 MHz in size}.

\subsection{Calibration and Imaging}
\label{section:cal_imaging}
We processed the data using the VLA pipeline 6.6.1\footnote{\url{https://science.nrao.edu/facilities/vla/data-processing/pipeline}} which is based on the Common Astronomy Software Applications \citep[CASA;][]{CASA2022}. The pipeline provided a series of automated recipes that flag, calibrate and image the data. Firstly, the data were exported from the VLA's standard archival format, a Science Data Model-Binary Data Format (SDM-SDF) file, to a measurement set. The measurement set was then Hanning smoothed to reduce Gibbs ringing, flagged, and then flux density scaled using a standard flux calibrator (3C48 for these data). Various calibration steps were then performed, including correcting for the gain curves, opacities, antenna position errors, requantisation biases, switched and system power corrections and finally delay and bandpass calibrations. Calibration tables generated during each of the steps were applied to the data. Additional flagging was performed, and science-ready calibrated visibilities were obtained.

To diagnose the quality of the calibration, we produced un-deconvolved images for each individual observing epoch of both datasets using the CASA task \texttt{tclean}.  We employed the multi-term multi-frequency \citep[MTMFS; ][]{Rau2011} \textcolor{black}{algorithm} to account for the large observing bandwidth (4 GHz). The images were Briggs weighted with a robust parameter of 0.5  \citep{Briggs1995}. The theoretical Stokes I thermal noise for 2M2228 in a dual polarisation Briggs weighted image at the robust parameter for each observation epoch was $\sim2.3\ \mu\text{Jy~beam}^{-1}$. Finally, we performed deep deconvolution for both Stokes parameters using \texttt{tclean} to suppress the sidelobes. We used the \texttt{auto-multithresh} algorithm \citep[][]{Kepley2020}, which is implemented in CASA, to perform automatic masking of emission during CLEAN \citep[][]{Hogbom1974,Clark1980}. We note that the Stokes V images are expected to be largely free of sidelobes as background radio sources, which are responsible for the sidelobes in the Stokes I, have low circular polarisation levels at $<1$\% \citep[e.g.][]{Macquart2003}.

\subsection{Astrometry}

The pointing centre of the array during the observations did not account for the proper motion of 2M2228. We used the All Wide-field Infrared Survey Explorer (AllWISE) catalogue \citep{Cutri2014}, a reprocessing of WISE \citep[][]{Wright2010} and identified the source at coordinates 22h28m29.0153s and $-$43d10m29.8312s in Right Ascension (RA) and Declination (Dec) respectively, at epoch 2010-07-23 (MJD 55400.0). The corresponding proper motions for the source are $102.3\pm5.8\ \text{mas yr}^{-1}$ and $-324.4\pm5.1\ \text{mas yr}^{-1}$ in RA and Dec \citep[][]{Faherty2012}.

\subsection{Radio Light Curves}

\label{section:variability}

To produce radio light curves, we closely followed the strategy of \citep[][]{Wandia2025} and employed \texttt{WSClean} \citep[][]{WSClean2014} to make a sky model by masking the target and deconvolving all other sources within the primary beam of the array. We then used the CASA task \texttt{ft} to add the modelled sources to the \texttt{modeldatacolumn} of the measurement set. The model was then subtracted from the visibilities using the task \texttt{uvsub} and the data phase shifted using the task \texttt{phaseshift} to the proper motion-corrected position of the target. We then accessed the visibilities stored in a measurement set using the \texttt{table} and \texttt{ms} tools available from the CASA toolkit \texttt{casatools}. Using these tools, we extracted and averaged the real parts of the visibilities over all baselines, channels and spectral windows for the parallel-hand  correlations, \texttt{RR} and \texttt{LL}. We then binned the visibilities at a cadence of two minutes. Next, we computed the Stokes I and V as the sum and difference of \texttt{RR} and \texttt{LL}, respectively.  We noted that flux scaling calibrators have associated errors due to difficulties in determining the true flux density. For C-band observations using 3C48 (0137+3309) for flux scaling, the error associated with the absolute flux density was$\sim10$\% \footnote{\url{https://www.vla.nrao.edu/astro/calib/vlacal/cal_mon/last/0137+3309.html}}. To estimate the $1\sigma$ uncertainties, we added the scaling error in quadrature with a thermal noise of $\sim16.7\ \mu\text{Jy beam}^{-1}$ associated with a binning cadence of 2 minutes. We also produced dynamic spectra by binning the visibilities of the phase centered, background subtracted source at a cadence equal to the integration time. The light curves and dynamic spectra were validated through comparison with results obtained using \texttt{DSTOOLS} \footnote{\url{https://github.com/askap-vast/dstools/tree/1e227ea26d1f6aed00afe56447ccbcdc61eae37a}}, a radio data processing tool that extracts and plots light curves and dynamic spectra from radio interferometer visibilities.

\section{Results and Discussion}


\textcolor{black}{Emission from 2M2228 is detected in Stokes I and V from both observing epochs. The flux densities were extracted from the images by fitting a Gaussian profile using the CASA tool \texttt{imfit} and are presented in Table~\ref{tab:stokes_summary}. The corresponding synthesised images reporting the detections are presented in Figure~\ref{fig:sources_2M228}. The emission peak is found at coordinates 22h28m29.0690s -43d10m30.8146s in the first epoch and at coordinates 22h28m29.0831s -43d10m31.1885s in the second epoch. Close examination of the peak positions of the emission shows a positional shift of $\sim405$ mas between the two observing epochs which is within the synthesised beam of $0.7$\arcsec . From the Stokes V measurements, the flux density changes sign from positive to negative between the two epochs, appearing to indicate \textcolor{black}{a reversal in the line-of-sight component of the magnetic field at the emitter as opposite poles rotate into view.}  
\textcolor{black}{We note the source appears resolved in the Stokes I images. The current data are insufficient to determine the cause. We will explore this further with future, higher sensitivity observations.} }

\begin{table}

    \newcolumntype{C}[1]{>{\centering\arraybackslash}p{#1}}
    \renewcommand{\arraystretch}{1.5}
    \centering
     \begin{tabularx}{\columnwidth}{>{\centering\arraybackslash}X|
                                 >{\centering\arraybackslash}X
                                 >{\centering\arraybackslash}X}
    
    \hline
    Epoch & Stokes & \makecell{Peak Flux Density \\ $(\mu\text{Jy beam}^{-1})$}  \\
    \hline
    1 & I & $\textcolor{black}{67.3\pm4.9\ \mu\ \rm{Jy beam}^{-1}}$ \\
     & V &  $\textcolor{black}{14.4\pm3.0\ \mu\text{Jy beam}^{-1}}$ \\
     \hline
    2 & I & $\textcolor{black}{107.2\pm5.2\ \mu\rm{Jy\ beam}^{-1}}$ \\
     & V & $\textcolor{black}{-20.7\pm1.2\ \mu\text{Jy beam}^{-1}}$ \\

    \hline
	\end{tabularx}
    \caption{\textcolor{black}{A summary of the measured Stokes I and V peak flux densities for each observing epoch.}}
    \label{tab:stokes_summary}
\end{table}

\begin{figure*}
  \centering
    \setlength{\tabcolsep}{-11pt} 
    \renewcommand{\arraystretch}{0} 
  \begin{tabular}{cccc}
      \includegraphics[width=0.6\columnwidth]{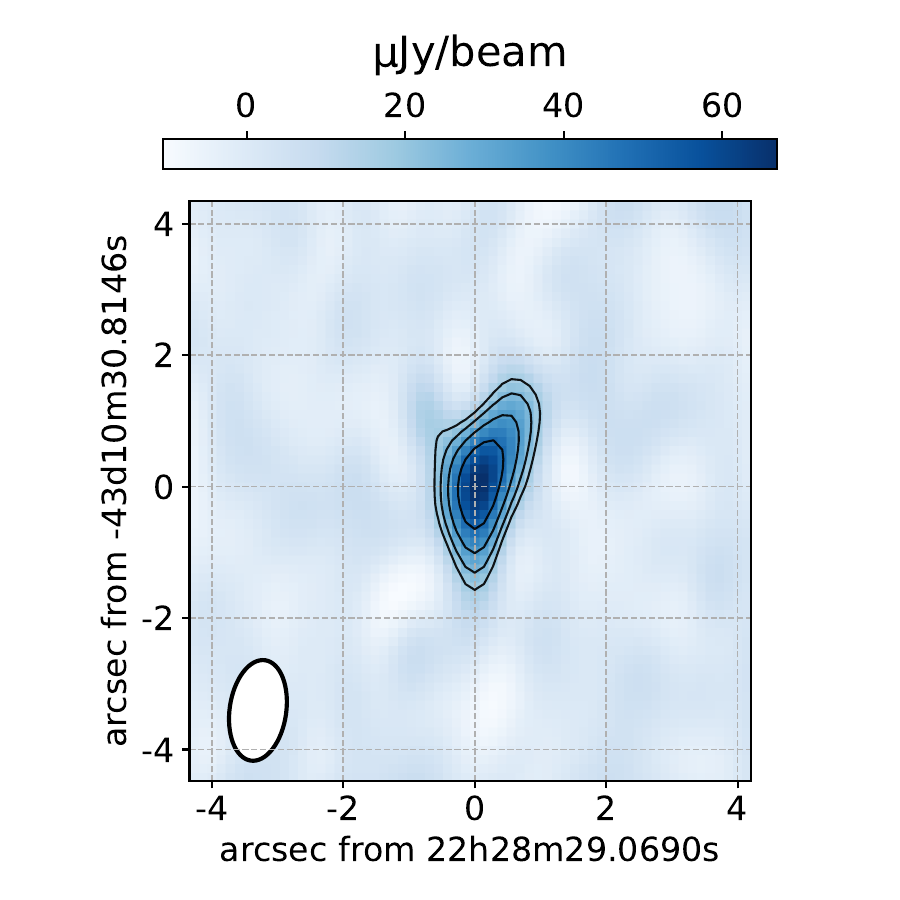}
      &
      \includegraphics[width=0.6\columnwidth]{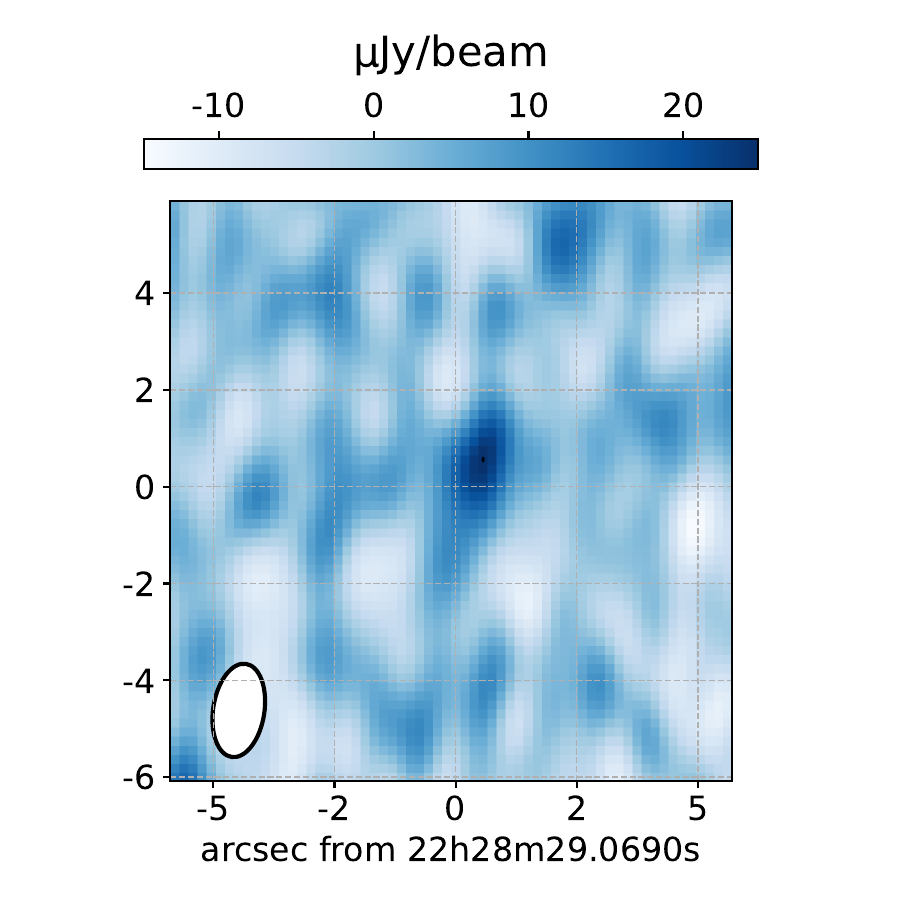}
      &
      \includegraphics[width=0.6 \columnwidth]{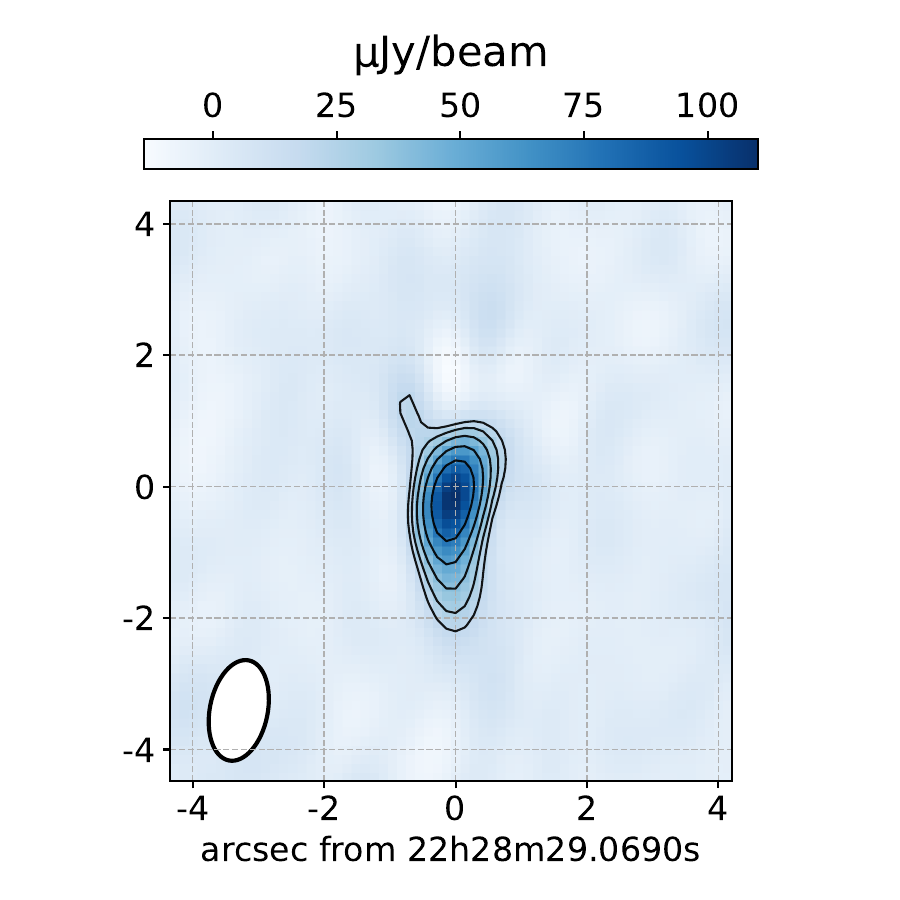}  
      
      &
      \includegraphics[width=0.6\columnwidth]{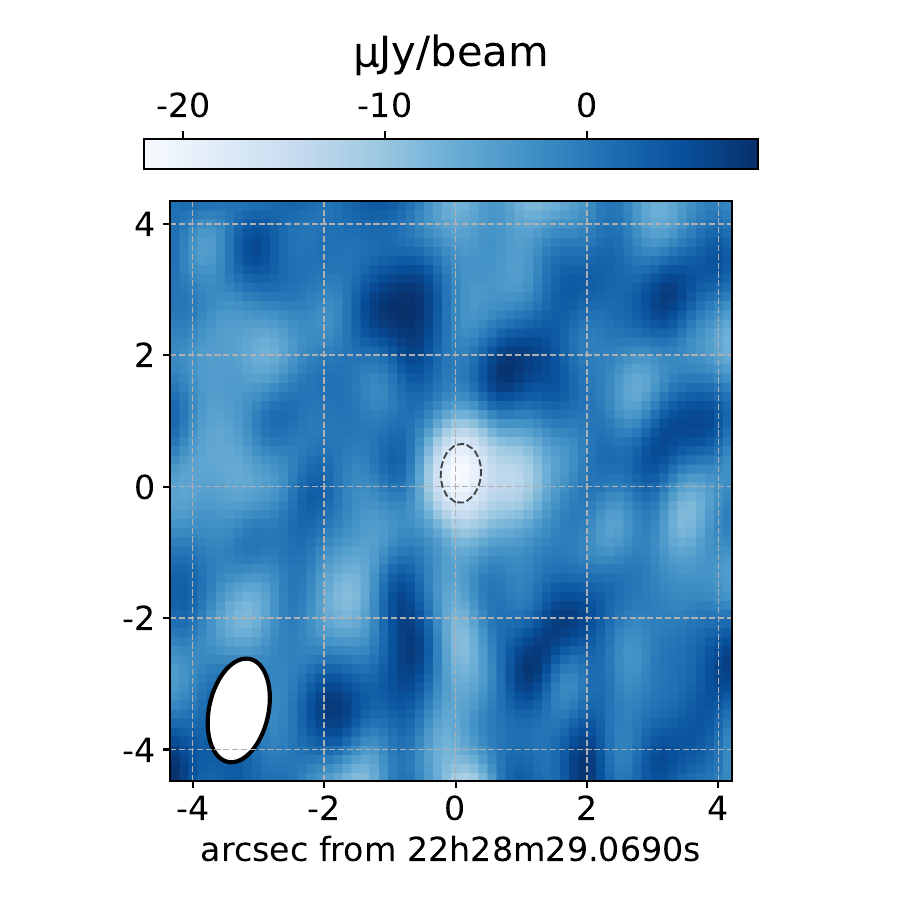} 
      \vspace{0.10cm}
      \\
      (a) & (b) & (c) & (d) 
      \vspace{0.10cm}
    \end{tabular}
    \caption{ Stokes I and V images of 2M2228 for the observations conducted over two epochs. Panels (a) and (b) correspond to observations of the first epoch, and panels (c) and (d) to epoch two observations. (a) The source is detected in the Stokes I at a peak flux density $\textcolor{black}{67.3\pm4.9\ \mu\text{Jy beam}^{-1}}$. The $1\sigma$ r.m.s. noise in the image is $3.1\ \mu\text{Jy\,beam}^{-1}$ resulting in a signal to noise ratio (SNR) $\sim22$. (b) The corresponding Stokes V image is detected at a peak flux density $\textcolor{black}{14.4\pm3.0\ \mu\text{Jy\,beam}^{-1}}$ at a $1\sigma$ r.m.s noise of $3.3\ \mu\text{Jy\,beam}^{-1}$ giving a SNR $\sim4$. (c) Stokes I image for the second observing epoch. The detection is at a peak flux density $\textcolor{black}{107.2\pm5.2\ \mu\text{Jy\,beam}^{-1}}$ at a $1\sigma$ r.m.s noise of $3.9\ \mu\text{Jy\,beam}^{-1}$ giving a SNR $\sim27$. (d) The corresponding Stokes V image for the observation.  The detection is at a peak flux density $\textcolor{black}{-20.7\pm1.2\ \mu\text{Jy\,beam}^{-1}}$ at a $1\sigma$ r.m.s noise of $3.5\ \mu\text{Jy\,beam}^{-1}$ giving a SNR $\sim6$. The peak flux is negative, indicating the left-hand circular polarisation is dominant. \textcolor{black}{All} the contours are drawn at $5\sigma\times(-4,-2\sqrt{2},-2,-\sqrt{2},-1,1)$. In all the images, positive contours are represented by the solid line and negative contours by the broken line. The  white ellipse with a black outline to the bottom left of the image represents the synthesised beam. All the positions are with respect to the detected positions at the first epoch. }
 \label{fig:sources_2M228}
\end{figure*}

\subsection{Temporal Variability}

\begin{figure*}
    \centering
    \subfloat[ ]{%
        \includegraphics[width=0.49\textwidth]{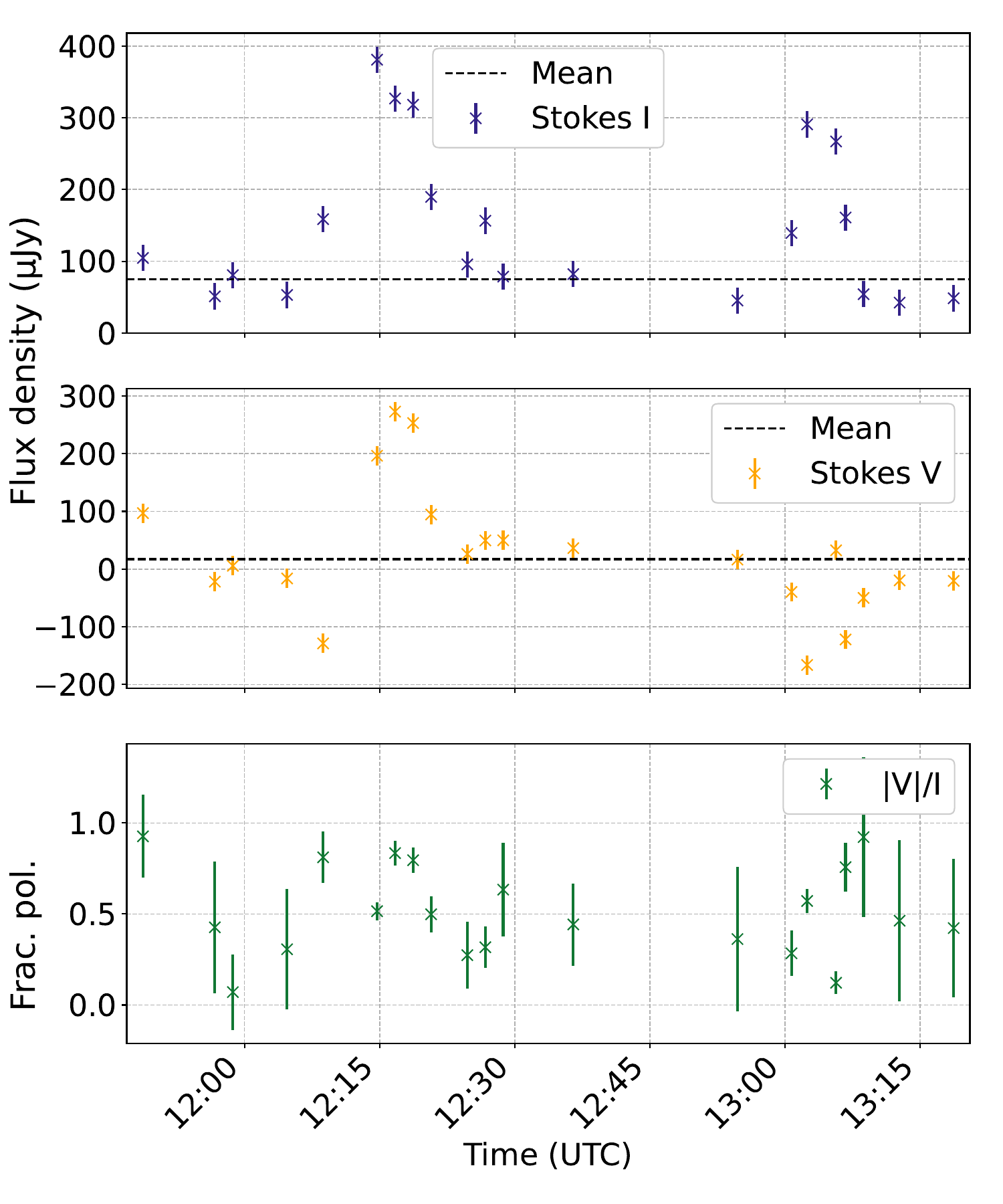}%
        \label{fig:sub1}
    }
    \hfill
    \subfloat[]{%
        \includegraphics[width=0.49\textwidth]{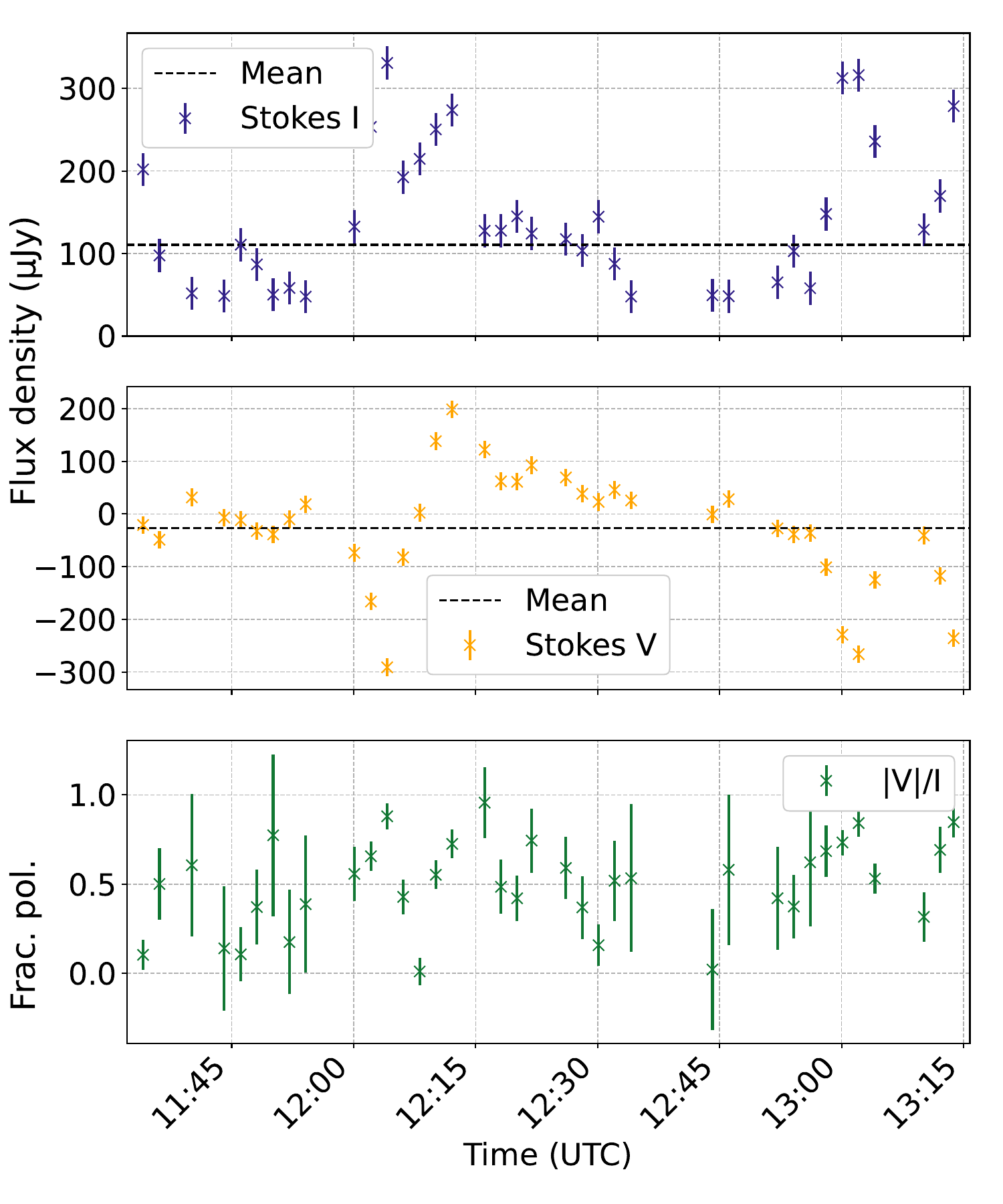}%
        \label{fig:sub2}
    }
    \caption{
    (a) Upper panel: Stokes I light curves for epoch one observations at 4-8 GHz with the data binned a cadence of 2 min. Center panel: the corresponding Stokes V light curve. The mean values for the Stokes I and V are $78.2\ \mu\text{Jy beam}^{-1}$ and $11.0\ \mu\text{Jy beam}^{-1}$. Lower panel: the fractional polarisation obtained as the ratio of the absolute Stokes V value to the Stokes I for each binned point.
    (b) Stokes I and V light curves for epoch two observations displayed using a similar panel layout to epoch one observations. The broken line represents the mean flux density. The mean values for the Stokes I and V are $122.1\ \mu\text{Jy beam}^{-1}$ and $-27.3\ \mu\text{Jy beam}^{-1}$. Lower panel: the fractional polarisation fraction obtained as the ratio of the absolute Stokes V value to the Stokes I for each binned point. The light curves were filtered to remove unphysical values, specifically cases with fractional polarization greater than 1 and data points with excessively large errors.
 }
    \label{fig:variability}
\end{figure*}

The light curves for 2M2228 presented in Figure~\ref{fig:variability} (a) and (b) for the Stokes I and V reveal the source is displaying temporal variability. In Table~\ref{tab:feature_summary} we present a summary of the timing, peak flux densities and their associated errors and fractional polarisation fractions of detected bursts. We note that all timestamps have been rounded to the nearest minute. Additionally, we present the fractional polarisation fractions in Figure~\ref{fig:variability} corresponding to each binned cadence for epochs one and two. Unfortunately, none of the bursts are captured in the dynamic spectra; consequently, the corresponding spectra are not presented.

We observe that in the bursts occur at an interval of $\sim47$ minutes in the first epoch and  $\sim58$ minutes in the second. 
Considering 2M2228 has a half period $42.9\pm4.8$ \textcolor{black}{minutes} \citep[][]{Clarke2008}, our analysis indicates the burst in the first epoch align within a half period timescale while the burst observed in the second epoch does not. We caution that the reported radio periodicities are tentative and precise periods can only be reliably determined through longer observations.  

\begin{table}
    \newcolumntype{C}[1]{>{\centering\arraybackslash}p{#1}}
    \renewcommand{\arraystretch}{1.2}
    \centering
    \begin{tabularx}{\columnwidth}{C{0.7cm}| C{0.7cm} C{0.7cm} C{2.1cm} C{0.7cm} C{0.8cm}}
    
    \hline
    Epoch & Burst & Stokes & \makecell{Peak Flux Density \\ $(\mu\text{Jy beam}^{-1})$} & Time (UTC) & $f_\text{c} (\%)$ \\
    \hline
    1 & \multirow{2}{*}{1} & I & 380.8$\pm18.3$& 12:15 &  \multirow{2}{*}{\centering $51.5\pm5.0$} \\
    &   & V & 196.3$\pm16.7$ & 12:15 & \\  \cline{2-6} 
    
    &  \multirow{2}{*}{2}  & I & 290.7$\pm20.0$ & 13:02 &  \multirow{2}{*}{\centering $57.2\pm7.0$}  \\
    &  & V & -166.2$\pm16.7$ & 13:02 & \\
    \hline
    \multirow{2}{*}{2} & \multirow{2}{*}{1} & I & 330.8$\pm18.3$ & 12:04 &  \multirow{2}{*}{\centering $88.0\pm7.0$}\\
    &  & V & -291.1$\pm16.7$ & 12:04 & \\ \cline{2-6} 
    
    & \multirow{2}{*}{2} & I & 316.1$\pm20.0$ & 13:02 &  \multirow{2}{*}{\centering $84.2\pm7.5$}\\
    &  & V & -266.1$\pm16.7$ & 13:02 & \\
    \hline
	\end{tabularx}
    \caption{A summary of the bursts observed from the light curve.}
    \label{tab:feature_summary}
\end{table}

\subsection{Emission Mechanism}

To characterise the radio emission, we first evaluate the brightness temperatures $T_\text{B}$ using equation~\ref{eqn:brightness_temp}
\begin{equation}
    T_\text{B} \simeq \frac{S_\nu}{1\ \mu\text{Jy}}\times\ \left(\frac{\nu}{\text{1 GHz}}\right)^{-2} \times \left(\frac{d}{\text{1 pc}}\right)^2\ \times \left(\frac{L}{\text{1 cm}}\right)^{-2} \times10^{26}\ \ \  \text{K} \ ,
    \label{eqn:brightness_temp}
\end{equation}
where $S_\nu$ is the flux density in $\mu\text{Jy beam}^{-1}$, $\nu$ is the central frequency of the observation (6 GHz), $d$ is the distance to the source in parsec, and $L$ is the size of the emitting region in cm. Assuming the source size is on the order of the radius of 2M2228, that is, $L\sim0.94\ R_\text{J}$ where \textcolor{black}{$R_\text{J}=7\times10^9$} cm is the Jovian radius, we determine $T_\text{B}\sim (4.9-7.8)\ \times10^{8}$ K for peak flux densities measured from the images made using time and frequency averaged Stokes I visibilities in epoch one and two, respectively. 
We infer \textcolor{black}{isotropic equivalent spectral luminosities} $L_\nu = 4\pi d^2S_\nu$ of $L_\nu\sim(9.1-14.5)\times10^{12}\ \text{erg s}^{-1}\text{Hz}^{-1}$ and radio luminosities $L_\text{R} = \int L_\nu\  \rm{d}\nu\sim(3.65-5.81)\times10^{22}\ \text{erg s}^{-1}$ for the first and second epochs, respectively over an observing bandwidth  d$\nu$.

To estimate the spectral indices, we divide the 4 GHz bandwidth into two equal sub-bands, each 2 GHz in size.  We then make images for each sub-band  and measure the integrated flux. We  estimate the spectral index $(S\propto\nu^{\alpha})$ $\alpha$ at $-2.56\pm0.61$ and $-1.65\pm0.40$ for the first and second epoch respectively. The steep spectral indices for the first epoch \textcolor{black}{may indicate} a cut-off frequency where the flux density rapidly diminishes and the source becomes undetectable.


Besides plasma emission, the electron cyclotron maser emission (ECME) has been shown to be the dominant mechanism for the generation of coherent radio emission in low mass stars and brown dwarfs at cm wavelengths \citep[][]{Hallinan2007,Hallinan2008}. ECME phenomenologically manifests as coherent, highly polarised bursts originating from spatially localised regions occurring at a local cyclotron frequency $\nu_\text{c} = 2.8 \times 10^{6}B$ Hz, where $B$ is the magnetic field strength in gauss, and exhibits a cut-off in frequency. Accordingly, we attribute the highly polarised emission observed in the light curves, which reaches $f_\text{c}\sim100\%$ to ECME, and constrain
the magnetic field strength of 2M2228 to $B\gtrsim 1.4$ kG.

\section{Summary and Future Work}

We have analysed archival VLA C band (4-8 GHz) band observations of the rapidly rotating T6/T6.5. brown dwarf 2MASS J22282889-4310262 (2M2228) over two observing epochs ($2\times96$ minutes) and detected bursts occuring at intervals of $\sim47$ and $\sim58$ minutes respectively. This detection makes 2M2228 the eighth and notably, the most rapidly rotating T dwarf detected at radio wavelengths. From our Stokes I images, we have estimated brightness temperatures and inferred \textcolor{black}{isotropic equivalent spectral luminosities} of $L_\nu\sim10^{12}\ \text{erg s}^{-1}\text{Hz}^{-1}$ and radio luminosities of $L_\text{R}\sim10^{22} \text{erg s}^{-1}$. The large fractional polarisation ratios measured, $f_\text{c}>50\%$, from the bursts suggest the emission is produced through the electron cyclotron maser emission (ECME) allowing us to constrain the magnetic field strength of 2M2228 to $B\gtrsim1.4$ kilogauss. 
We caution that the inferred periodicities are provisional considering the brief observation durations.

\textcolor{black}{The VLA detection of periodic, highly polarised radio bursts from 2M2228, combined with the demonstrated long-term stability of its atmospheric structure \citep[][]{Yang2016}, makes this brown dwarf an excellent target for coordinated multi-wavelength observation. Joint radio and James Webb Space Telescope (JWST) monitoring could provide complementary insights, enhancing the prospects of detecting sustained auroral emission. The persistence of the phase shifted between different wavelengths in the mid infrared light curves over thousands of rotations indicates that the underlying atmospheric structure maybe stable on multi-year timescales, increasing the likelihood that auroral phenomena are sustained rather than transient. JWST's unprecedented sensitivity and spectral resolutions in the near and mid infrared may enable the detection of faint auroral tracers such as H$^+_3$ that were previously inaccessible to other facilities \citep[][]{Gibbs2022,Pineda2024}. Moreover, its ability to perform time-resolved spectroscopy could distinguish auroral emission from associated weather phenomena through its characteristic spectral signatures. Recent work has demonstrated the instruments ability to map atmospheric features in the planetary mass radio loud T dwarf SIMP J0136+0933 \citep[][]{McCarthy2025}, highlighting the potential for infrared auroral emission to contribute to complex atmospheric features.}

\textcolor{black}{Together, the combination of atmospheric stability and detected radio emission makes 2M2228 a particularly promising laboratory for testing the framework proposed by \cite{Hallinan2015}, in which large magnetospheric currents drive multi-wavelength aurorae and influence atmospheric weather. More generally such observations will advance our understanding of auroral processes and their connection to atmospheric dynamics in T-type brown dwarfs, which serve as analogs to many directly imaged gas giant planets.}



\section*{Acknowledgements}

This project has been made possible in part by a grant from the SETI Institute. This work made use of Astropy:3, a community-developed core Python package and an
ecosystem of tools and resources for astronomy \citep[][]{astropy2013,astropy2018,astropy2022}. The National Radio Astronomy Observatory is a facility of the National Science Foundation
operated under cooperative agreement by Associated Universities, Inc. This research has made use of the SIMBAD database, operated at CDS, Strasbourg, France. This research has made use of the VizieR catalogue access tool, CDS, Strasbourg, France. This publication makes use of data products from the Wide-field Infrared Survey Explorer, which is a joint project of the University of California, Los Angeles, and the Jet Propulsion Laboratory/California Institute of Technology, funded by the National Aeronautics and Space Administration.

\section*{Data Availability}

Data underlying this article are publicly available in the NRAO Data Archive at \url{https://data.nrao.edu/portal} and can be accessed with project code 15A-045.



\bibliographystyle{mnras}
\bibliography{example} 








\bsp	
\label{lastpage}
\end{document}